\documentclass[a4paper,superscriptaddress,floatfix,eqsecnum,aps,nofootinbib]{revtex4}
\usepackage[dvips]{epsfig}
\usepackage{amsmath}
\usepackage{enumerate}

\newcommand{\fourgraphs}[4]{%
\unitlength=1in
\begin{picture}(5.8,5.2)(-0.15,0)
\put(0,0){\epsfig{file=#3, width=\wth}}
\put(2.7,0){\epsfig{file=#4, width=\wth}}
\put(0,2.6){\epsfig{file=#1, width=\wth}}
\put(2.7,2.6){\epsfig{file=#2, width=\wth}}
\end{picture}}

\newlength{\wth}
\newcommand{\sixgraphs}[6]{%
\unitlength=1in
\begin{picture}(8.7,7.1)(-0.7,-0.1)
\put(-0.5,4.8){\epsfig{file=#1, width=\wth}}
\put(2.7,4.8){\epsfig{file=#2, width=\wth}}
\put(-0.5,2.3){\epsfig{file=#3, width=\wth}}
\put(2.7,2.3){\epsfig{file=#4, width=\wth}}
\put(-0.5,-0.2){\epsfig{file=#5, width=\wth}}
\put(2.7,-0.2){\epsfig{file=#6, width=\wth}}
\end{picture}}

\newcommand{\GeV}{\, \text{GeV}}

\begin{document}

\title{Stochastic superspace phenomenology at the Large Hadron Collider}
\author{Archil Kobakhidze}
\email{archilk@unimelb.edu.au}
\affiliation{ARC Centre of Excellence for Particle Physics at the Terascale, School of Physics, The University of Melbourne, Victoria 3010, Australia}
\author{Nadine Pesor}
\email{npesor@student.unimelb.edu.au}
\affiliation{ARC Centre of Excellence for Particle Physics at the Terascale, School of Physics, The University of Melbourne, Victoria 3010, Australia}
\author{Raymond R. Volkas}
\email{raymondv@unimelb.edu.au}
\affiliation{ARC Centre of Excellence for Particle Physics at the Terascale, School of Physics, The University of Melbourne, Victoria 3010, Australia}
\author{Martin J. White}
\email{mwhi@unimelb.edu.au}
\affiliation{ARC Centre of Excellence for Particle Physics at the Terascale, School of Physics, The University of Melbourne, Victoria 3010, Australia}

\begin{abstract}
We analyse restrictions on the stochastic superspace parameter space arising from 1 fb$^{-1}$ of LHC data, and bounds on sparticle masses, cold dark matter relic density and the branching ratio of the process $B_s \rightarrow \mu^+ \mu^-$. A region of parameter space consistent with these limits is found where the stochasticity parameter, $\xi$, takes values in the range $-2200 \GeV < \xi < -900 \GeV$, provided the cutoff scale is $\mathcal{O}(10^{18}) \GeV$.
\end{abstract}

\maketitle

Supersymmetry is a popular extension to the Standard Model, renowned for its solution to the hierarchy problem, while also providing a dynamical mechanism of electroweak symmetry breaking and a dark matter candidate in its minimal form (see  \cite{Martin:1997ns} for a review). By postulating a symmetry relating fermionic and bosonic states, it predicts the existence of a superpartner of opposite spin statistics and degenerate mass for each Standard Model particle. These superpartners have not been observed, requiring that supersymmetry be softly broken for it to remain a viable model.

A number of methods for breaking supersymmetry exist, which can be broadly categorised into two groups: those originating from a fundamental theory where supersymmetry is spontaneously broken in a hidden sector, then communicated to the visible sector by a messenger, and the purely phenomenological approach, where no explanation is provided as to the origins of supersymmetry breaking. Although fundamental theories provide a dynamical explanation for the mechanism of supersymmetry breaking, complications such as large sparticle induced FCNC amplitudes in gravity mediation, and CP violation issues in gauge mediation, leave no single compelling solution \cite{Kolda:1997wt}. Alternatively, the phenomenological approach of writing down the most general Lagrangian with explicit soft-breaking terms fails to address the source of SUSY breaking, while introducing an overwhelmingly vast new parameter space. 

Stochastic superspace is a unique mechanism for softly breaking supersymmetry that bridges the two approaches discussed above \cite{Kobakhidze:2008py}. By considering the Grassmannian coordinates to be stochastic variables, a very constrained set of soft-breaking terms emerges. Though the underlying cause of stochasticity is not postulated, it introduces predictability not achievable through the purely phenomenological approach.

In this paper we investigate the phenomenology of stochastic superspace at the LHC. For each suitable point in the stochastic superspace parameter space, we use a fast simulation of the ATLAS detector to investigate whether points are excluded by the ATLAS zero lepton searches, which in turn are amongst the most constraining current limits on direct sparticle production. 

The rest of the paper is structured as follows: Sec. \ref{sec:overview} provides an overview of stochastic superspace, Sec. \ref{sec:scan} discusses constraints applied in a scan of its parameter space, and Sec. \ref{sec:LHC} describes the technique used to simulate stochastic superspace model points at the ATLAS detector, and explains how we approximate the ATLAS exclusion limits on supersymmetric particle production. We terminate with concluding remarks in Sec. \ref{sec:conclusion}.

\section{Overview of stochastic superspace}
\label{sec:overview}

The basis of stochastic superspace models is the assumption that the Grassmannian coordinates, $\theta$ and $\bar{\theta}$, are stochastic variables. Writing down the most general probability distribution consistent with Lorentz invariance, we find that only one additional parameter is required such that \cite{Kobakhidze:2008py},
\begin{equation}
\label{eq:probDist}
 \mathcal{P} \left(\theta, \bar{\theta} \right) \left| \xi \right|^2 = 1 + \xi^* \left( \theta \theta \right) +  \xi \left( \bar{\theta} \bar{\theta} \right) + \left| \xi \right|^2 \left( \theta \theta \right)  \left( \bar{\theta} \bar{\theta} \right),
\end{equation}
where $\xi$, a complex number of mass dimension, is the measure of stochasticity. The Lagrangian in ordinary spacetime is then found by averaging the supersymmetric Lagrangian over the probability distribution in Eq. \eqref{eq:probDist}. Applying this procedure to the superpotential of the minimal supersymmetric standard model,
\begin{equation}
 W_{\text{SM}} = \mu H_u H_d + \hat{y}^{\text{up}} Q U^c H_u + \hat{y}^{\text{down}} Q D^c H_d + \hat{y}^{\text{lept}} L E^c H_d,
\end{equation}
leads to the soft-breaking terms
\begin{equation}
\label{eq:sslagrangian}
 L_{\text{soft scalar}} = -\xi^* \mu \tilde{H}_u \tilde{H}_d - 2 \xi^* \left[ \hat{y}^{\text{up}} \tilde{Q} \tilde{U}^c \tilde{H}_u + \hat{y}^{\text{down}} \tilde{Q} \tilde{D}^c \tilde{H}_d + \hat{y}^{\text{lept}} \tilde{L} \tilde{E}^c \tilde{H}_d \right] + \text{h.c.},
\end{equation}
where tildes represent the scalar component of the chiral superfield. Similarly, averaging over the gauge-kinetic $F$ densities results in soft-breaking masses for the gauginos, $\lambda^{\left( i \right)}\left(x\right)$,
\begin{equation}
\label{eq:gaugelagrangian}
 L_{\text{gauge}} = \left[ \frac{1}{2} \sum_i \text{Tr} W^{\left(i\right) \alpha} W_{\alpha}^{\left(i\right)} \right]_F - \frac{\xi^*}{2} \sum_i \text{Tr} \lambda^{\left(i\right)} \lambda^{\left(i\right)} + \text{h.c.},
\end{equation}
where $W^{\left(i\right) \alpha}$ denotes the standard model field-strength superfields. See \cite{Kobakhidze:2008py} for further details. Referring to Eqns. \eqref{eq:sslagrangian} and \eqref{eq:gaugelagrangian}, it is clear that the tree-level soft-breaking terms for minimal stochastic superspace are \cite{Kobakhidze:2008py}: 
\begin{enumerate}[(i)]
 \item the bilinear Higgs soft term, $B_{\mu} = \xi^{*}$,
 \item the universal trilinear scalar soft terms proportional to the Yukawa couplings, $A_0 = 2 \xi^{*}$,
 \item the universal gaugino masses, $m_{1/2} = \frac{1}{2} \left| \xi \right|$,
 \item the universal soft scalar masses, $m_0^2 = 0$.
\end{enumerate}
As these soft-breaking terms are renormalised at the quantum level, the relations listed above are only defined as such at some energy cutoff scale, $\Lambda$, which becomes an additional parameter of the theory.

The emergence of a pattern of soft-breaking terms dependent only on the parameter $\xi$ categorises stochastic superspace models as a special case of the broader constrained supersymmetric standard model (CMSSM). Although models with absent soft scalar masses are typically considered excluded\footnote{This is because under the assumption of R-parity conservation the lightest stau is a stable lightest supersymmetric particle (LSP) with the mass below the CDF model-independent limit on the mass of a charged massive stable particle (CHAMP), $m_{\rm CHAMP}\gtrsim 250 \GeV$ at 95$\%$ C.L. \cite{Aaltonen:2009kea}}, this is because the soft-breaking parameters are customarily defined at $\Lambda = M_{\text{GUT}}\approx 10^{16} \GeV$. We have found that for larger cutoff scales, $\Lambda \gtrsim 10^{18} \GeV$, the neutralino becomes the LSP, and consequently, one uncovers a new phenomenologically viable region of parameter space \cite{Kobakhidze:2008py,Kobakhidze:2010ee}. We look at this in further detail in Sec. \ref{sec:scan} by computing the neutralino relic density.

\section{Scan of parameter space}
\label{sec:scan}

We have previously demonstrated the existence of phenomenologically viable regions of the $\left( \xi, \Lambda \right)$ parameter space falling within the stau coannihilation regime \cite{Kobakhidze:2008py,Kobakhidze:2010ee}. In this study we expand the scope of the analysis primarily to examine a broader range of $\xi$ values and include a calculation of the cold dark matter relic density.

Most publicly available sparticle spectrum software packages take the mSUGRA parameters $m_0$, $m_{1/2}$, $A_0$, $\tan\beta$ and $\text{sgn}\mu$ as inputs at the cutoff scale. However, stochastic superspace makes a prediction for the bilinear Higgs soft term $B_{\mu}$, rather than explicitly predicting a value for $\tan\beta$. To facilitate a more thorough analysis of the model, we previously modified the program {\tt SOFTSUSY 3.1.7} \cite{Allanach:2001kg} to swap $\tan\beta$ for $B_{\mu}$ as an input. As branching ratios are required for this analysis, we use the package {\tt Isajet 7.81} \cite{Paige:2003mg} to calculate the sparticle and decay spectra, using the value of $\tan\beta$ found with {\tt SOFTSUSY}. Various packages included in {\tt IsaTools} calculate the relic density and branching ratios of rare decays.

We scan the parameter regions
\begin{align}
 -2600 \GeV \leq \xi \leq -200 \GeV, \\
 M_{\text{GUT}} \leq \Lambda \leq M_{\text{Pl}},
\end{align}
over 150 points in $\xi$ and 50 points in $\Lambda$ for a total of 7500 points, against the following conditions:
\begin{align}
 m_{\tilde{\chi}_1^0} < m_{\tilde{\tau}_1}, \label{eq:lsp} \\
 m_h > 109 \GeV, \label{eq:higgs}\\
 m_{\tilde{t}_R} > 95.7 \GeV, \label{eq:stop}\\
 m_{\tilde{b}_R} > 89 \, \GeV, \label{eq:sbot}\\
 m_{\tilde{\tau}_R} > 81.9 \GeV, \label{eq:stau}\\
 \text{BR}\left(B_s \rightarrow \mu^+ \mu^- \right) < 1.2 \times 10^{-8}, \label{eq:bll}\\
 0.0988 < \Omega h^2 < 0.1252. \label{eq:rd}
\end{align}
In stochastic superspace, the character of the LSP is dependent on one's parameter choice, and will either be the lightest neutralino or the lightest stau. Since R-parity is assumed to be conserved, we enforce the condition in Eq. \eqref{eq:lsp} to ensure all instances of a stau LSP are excluded. The lightest $CP-$even Higgs must satisfy condition \eqref{eq:higgs}, where we take into consideration the theoretical error of $3-5 \GeV$  \cite{Allanach:2004rh} in the published limit of $m_h > 114.4 \GeV$ at 95\% CL  \cite{Nakamura:2010zzi}. Conditions \eqref{eq:stop}-\eqref{eq:stau} arise from supersymmetric particle searches \cite{Nakamura:2010zzi}. Condition \eqref{eq:bll}, the upper limit on the branching fraction of the process $B_s \rightarrow \mu^+ \mu^-$ is as measured by the CMS and LHCb collaborations \cite{Chatrchyan:2011kr,Serrano:2011px}. Finally, condition \eqref{eq:rd} shows the acceptable range of relic density based on the WMAP 7-year mean, $\Omega h^2 = 0.1120 \pm 0.0056$ \cite{Komatsu:2010fb}, and the theoretical error of $\pm 0.012_{\text{SUSY}}$ \cite{Buchmueller:2011sw}, shown with errors added in quadrature. 

\begin{figure}
 \begin{center}
 \includegraphics{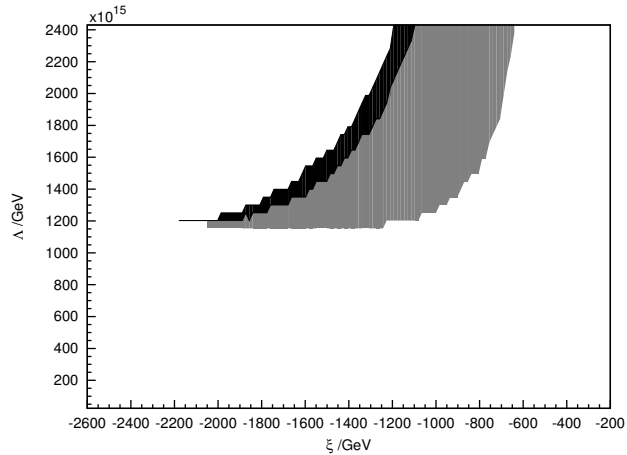}
 \caption{The phenomenologically viable region of parameter space, where the black area shows points that satisfy the WMAP bounds on relic density, and the gray area shows points where the relic density is lower than observed, but satisfies all other constraints.}
 \label{fig:rd}
 \end{center}
\end{figure}

\begin{figure}
 \begin{center}
 \includegraphics{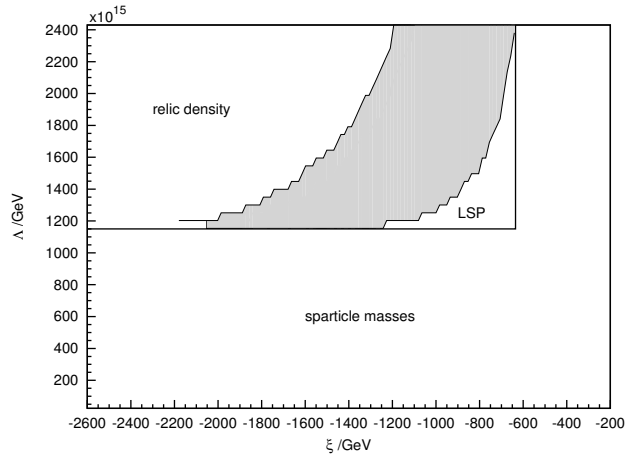}
 \caption{The causes of data point exclusion in the parameter space scan is shown in this plot. The viable region shown in gray is surrounded by three types of exclusion zones: too small or tachyonic sparticle masses are responsible for the exclusion of points in the region labelled ``sparticle masses,'' relic density higher than the upper WMAP bound excludes points in the region labelled ``relic density,'' and a stau LSP is responsible for excluding points in the region labelled ``LSP.'' }
 \label{fig:rdexcl}
 \end{center}
\end{figure}

Applying all constraints listed in Eqns. \eqref{eq:lsp}-\eqref{eq:rd}, we find a phenomenologically viable region of parameter space, shown in black in Fig. \ref{fig:rd}. 
Relaxing condition \eqref{eq:rd}, we find another region of interest where $\Omega h^2 \leq 0.0988$ (see gray region in Fig. \ref{fig:rd}). Although the relic density is lower than observed, this region is not necessarily excluded as the minimal model of stochastic superspace may be extended to include an additional source of dark matter. Furthermore, theoretical uncertainties in the determination of the sparticle mass spectrum can have a significant effect on the calculated value of the relic density, where a 1\% difference in mass could affect the calculated relic density by up to 10\% \cite{Belanger:2005jk,Allanach:2004xn}. It is possible, therefore, that the region of parameter space satisfying the relic density experimental bounds may be larger than depicted in Fig \ref{fig:rd}. The region where $\Lambda < 1.15 \times 10^{18} \GeV$ and/or $\xi > -600 \GeV$ is excluded due to sparticle masses violating conditions \eqref{eq:stop}-\eqref{eq:stau} or going tachyonic. Excluded points in the region where $\xi < -1200 \GeV$ and $\Lambda > 1.15 \times 10^{18} \GeV$ have relic density larger than the upper bound in condition \eqref{eq:rd}. In the region with $-1200 \GeV < \xi < -640 \GeV$ and $\Lambda > 1.15 \times 10^{18} \GeV$, a number of points are excluded for having a stau LSP. For clarity, Figure \ref{fig:rdexcl} displays these regions graphically.
All points in the parameter space satisfying constraints \eqref{eq:lsp}-\eqref{eq:bll} above, with relic density within or below the bound given in condition \eqref{eq:rd}, form the phenomenologically viable region of stochastic superspace.

\section{Impact of LHC exclusion limits}
\label{sec:LHC}
\subsection{LHC SUSY searches}
The ATLAS and CMS experiments~\cite{Aad:2008zzm,cms:2008zzk} have recently updated their searches for supersymmetric particles using the 2011 dataset~\cite{Aad:2011ib,Aad:2011zj,Aad:2011cw,Aad:2011qa,ATLAS:2011ad,Chatrchyan:2011zy,CMS-PAS-SUS-11-005,CMS-PAS-SUS-11-004}. Data collected from proton collisions at the Large Hadron Collider at $\sqrt{s}=7$~TeV are analysed in a variety of final states. No excess over the standard model expectation has been observed, allowing one to set exclusion limits in a variety of candidate model spaces such as the CMSSM, the model that most closely resembles that considered here. As the LHC is a proton-proton collider, one expects to dominantly produce coloured objects such as squarks and gluinos, whose inclusive leptonic branching ratios are relatively small, and hence the strongest CMSSM exclusions to date result from the ATLAS searches for events with no leptons and the CMS searches for sparticle production in hadronic final states. The ATLAS and CMS limits have a similar reach in the squark and gluino masses, and here we consider only the ATLAS zero lepton limits for simplicity. Recent interpretations of LHC limit results can be found in~\cite{Buchmueller:2011aa,Desai:2011th,Buchmueller:2011sw,Beskidt:2011qf,Buchmueller:2011ki,Allanach:2011qr}.

ATLAS defined a series of signal regions in which to look for sparticle production, each tuned to enhance sensitivity in a particular region of the $m_0$-$m_{1/2}$ plane. Events with an electron or muon with $p_T >$ 20 GeV were rejected. Table~\ref{tab:cuts} summarises the remaining selection cuts for each region, whilst Table~\ref{tab:results} gives the observed and expected numbers of events. These numbers were used by the ATLAS collaboration to derive limits on $\sigma \times A \times \epsilon$, where $\sigma$ is the cross-section for new physics processes for which the ATLAS detector has an acceptance $A$ and a detector efficiency of $\epsilon$. These results are also quoted in Table~\ref{tab:results}.

\begin{table}
\begin{center}
\begin{tabular}{|c|c|c|c|c|c|}
\hline
Region & R1 & R2 & R3 & R4 & RHM\\
\hline
Number of jets&$\ge 2$&$\ge 3$&$\ge 4$&$\ge 4$&$\ge 4$\\
$E_T^\mathrm{miss}$ (GeV) & $>130$&$>130$ &$>130$ &$>130$ &$>130$ \\
Leading jet $p_{T}$ (GeV) & $>130$  &$>130$ &$>130$ &$>130$ &$>130$ \\
Second jet $p_{T}$ (GeV)& $>40$  & $>40$  &$>40$  & $>40$  & $>80$  \\
Third jet $p_{T}$ (GeV)& - & $>40$  &$>40$  & $>40$  & $>80$  \\
Fourth jet $p_{T}$ (GeV) & - & -  &$>40$  & $>40$  & $>80$  \\
$\Delta\phi($jet, $p_{T}^\mathrm{miss})_\mathrm{min}$ & $>0.3$ & $>0.25$ & $>0.25$ & $>0.25$ & $>0.2$\\
$m_\mathrm{eff} (GeV)$ & $>1000$ &  $>1000$  &  $>500$  &  $>1000$  &  $>1100$ \\
\hline
\end{tabular}
\end{center}
\caption{\label{tab:cuts} Selection cuts for the five ATLAS zero lepton signal regions. $\Delta\phi($jet, $p_{T}^\mathrm{miss})_\mathrm{min}$ is the smallest of the azimuthal separations between the missing momentum $p_{T}^\mathrm{miss}$ and the momenta of jets with $p_T>$ 40 GeV (up to a maximum of three in descending $p_T$ order). The effective mass $m_\mathrm{eff}$ is the scalar sum of $E_T^\mathrm{miss}$ and the magnitudes of the transverse momenta of the two, three and four highest $p_T$ jets depending on the signal region. In the region RHM, all jets with $p_T>$40 GeV are used to define $m_\mathrm{eff}$.}
\end{table}

\begin{table}
\begin{center}
\begin{tabular}{|c|c|c|c|c|c|}
\hline
Region & R1 & R2 & R3 & R4 & RHM\\
\hline
Observed & 58 & 59 & 1118 & 40 & 18\\
Background & $62.4\pm 4.4\pm 9.3$&$54.9\pm 3.9\pm 7.1$&$1015 \pm 41\pm144$&$33.9\pm2.9\pm6.2$&$13.1\pm1.9\pm2.5$\\
$\sigma \times A \times \epsilon$ (fb)&22&25&429&27&17\\
\hline
\end{tabular}
\end{center}
\caption{\label{tab:results} Expected background yields and observed signal yields from the ATLAS zero lepton search using 1 fb$^{-1}$ of data~\cite{Aad:2011ib}. The final row shows the ATLAS limits on the product of the cross-section, acceptance and efficiency for new physics processes.}
\end{table}

The ATLAS collaboration has used the absence of evidence of sparticle production in 1 fb$^{-1}$ of data to place an exclusion limit at the 95\% confidence level in the $m_0$-$m_{1/2}$ plane of the CMSSM for fixed $A_0$ and tan$\beta$, and for $\mu>0$. Although the stochastic SUSY model considered here may be considered a subset of the CMSSM, it is non-trivial to recast this limit into a constraint on the parameters $\xi$ and $\Lambda$. Given a signal expectation for a particular model, however, one can easily evaluate the likelihood of that model using the published ATLAS background expectation and observed event yield in each search channel. By simulating points in the $\xi$-$\Lambda$ plane, we can therefore investigate the LHC exclusion reach in stochastic SUSY space, provided that we can demonstrate that our simulation provides an adequate description of the ATLAS detector.

\subsection{Simulation details and validation}
In this paper, we use \tt Isajet 7.81 \rm\cite{Paige:2003mg} to produce SUSY mass and decay spectra then use \tt Herwig++ 2.5.1 \rm\cite{Gieseke:2011na} to generate 10,000 Monte Carlo events. \tt Delphes 1.9 \rm\cite{Ovyn:2009tx} is subsequently used to provide a fast simulation of the ATLAS detector. The total SUSY production cross-section is calculated at next-to-leading order using \tt PROSPINO 2.1 \rm\cite{Beenakker:1996ed}, where we include all processes except direct production of neutralinos, charginos and sleptons since the latter are subdominant. The ATLAS setup differs from this only in the final step of detector simulation, where a full, \tt GEANT 4 \rm based simulation \cite{Agostinelli:2002hh} is used to provide a very detailed description of particle interactions in the ATLAS detector at vast computational expense. 

It is clear that the \tt Delphes \rm simulation will not reproduce every result of the advanced simulation. Nevertheless, one can assess the adequacy of our approximate results by trying to reproduce the ATLAS CMSSM exclusion limits. We have generated a grid of points in the $m_0$-$m_{1/2}$ plane using the same fixed values of tan$\beta$=10 and $A_0$=0 as the published ATLAS result. We must now choose a procedure to approximate the ATLAS limit setting procedure. ATLAS use both $CL_s$ and profile likelihood methods to obtain a 95\% confidence limit, using a full knowledge of the systematic errors on signal and background. Although the systematic error on the background is provided in the ATLAS paper, we do not have full knowledge of the systematics on the signal expectation, which may in general vary from point to point in the $m_0$-$m_{1/2}$ plane. Rather than implement these statistical techniques, we take a similar approach to that used in~\cite{Allanach:2011qr}, and use the published $\sigma \times A \times \epsilon$ limits to determine whether a given model point is excluded in a search channel. We use our simulation to obtain the $\sigma \times A \times \epsilon$ value for a given model point, and consider the model to be excluded if the value lies above the limit given in Table~\ref{tab:results}. This allows us to draw an exclusion contour in each search channel, and we estimate the combined limit by taking the union of the individual exclusion contours for each channel (i.e. the most stringent search channel for a given model is used to determine whether it is excluded). 

\begin{figure}
\sixgraphs{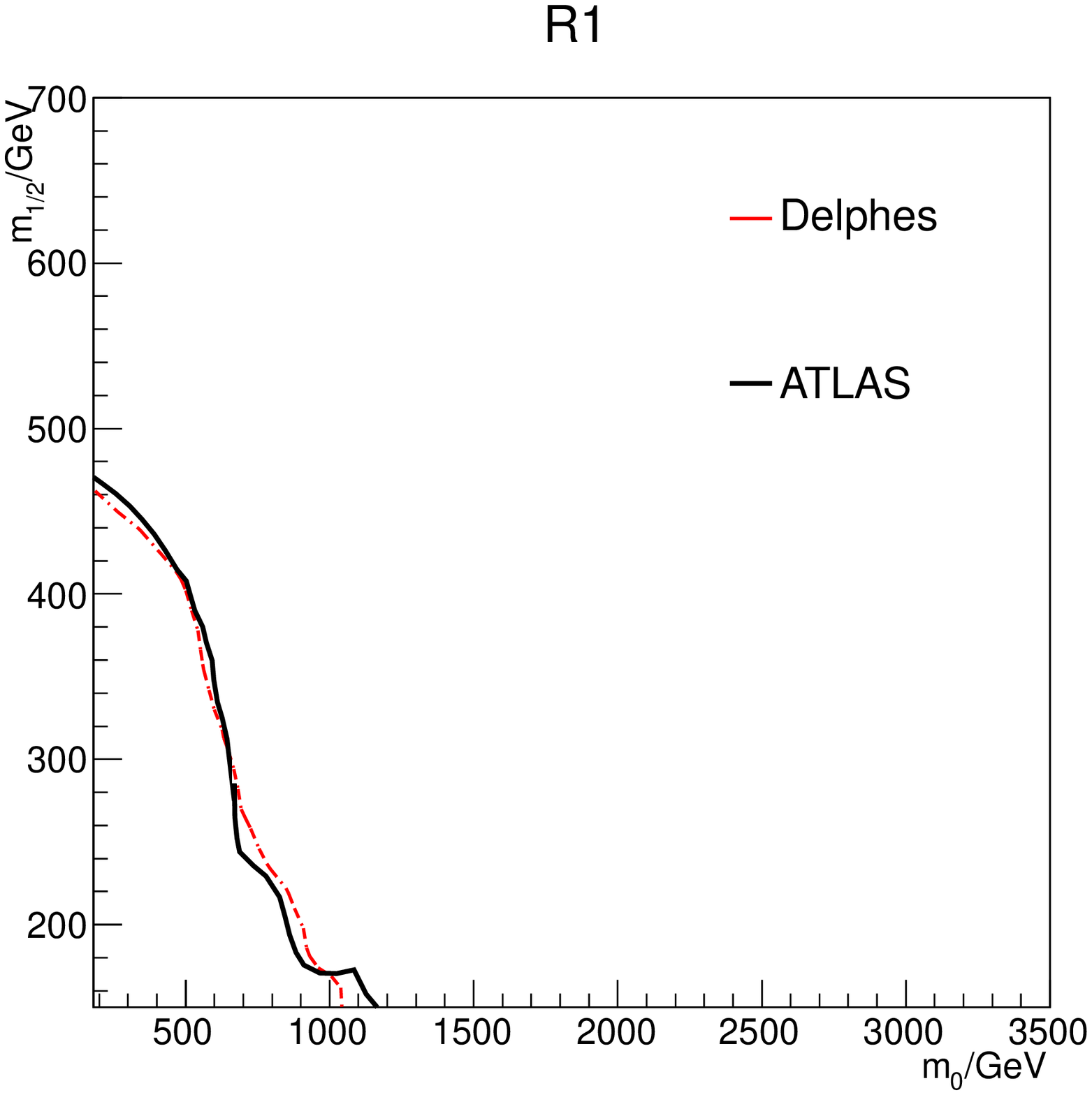}{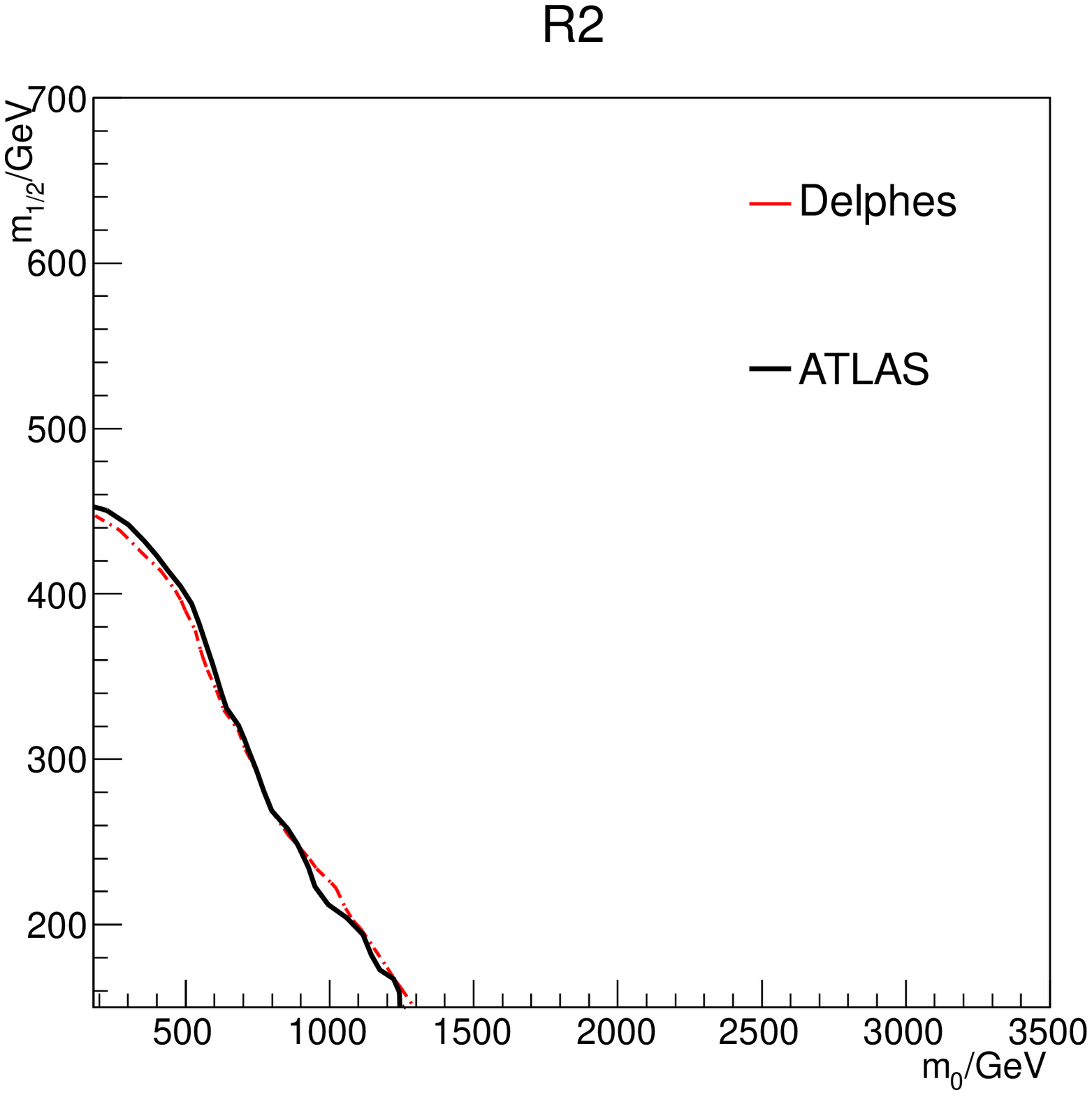}{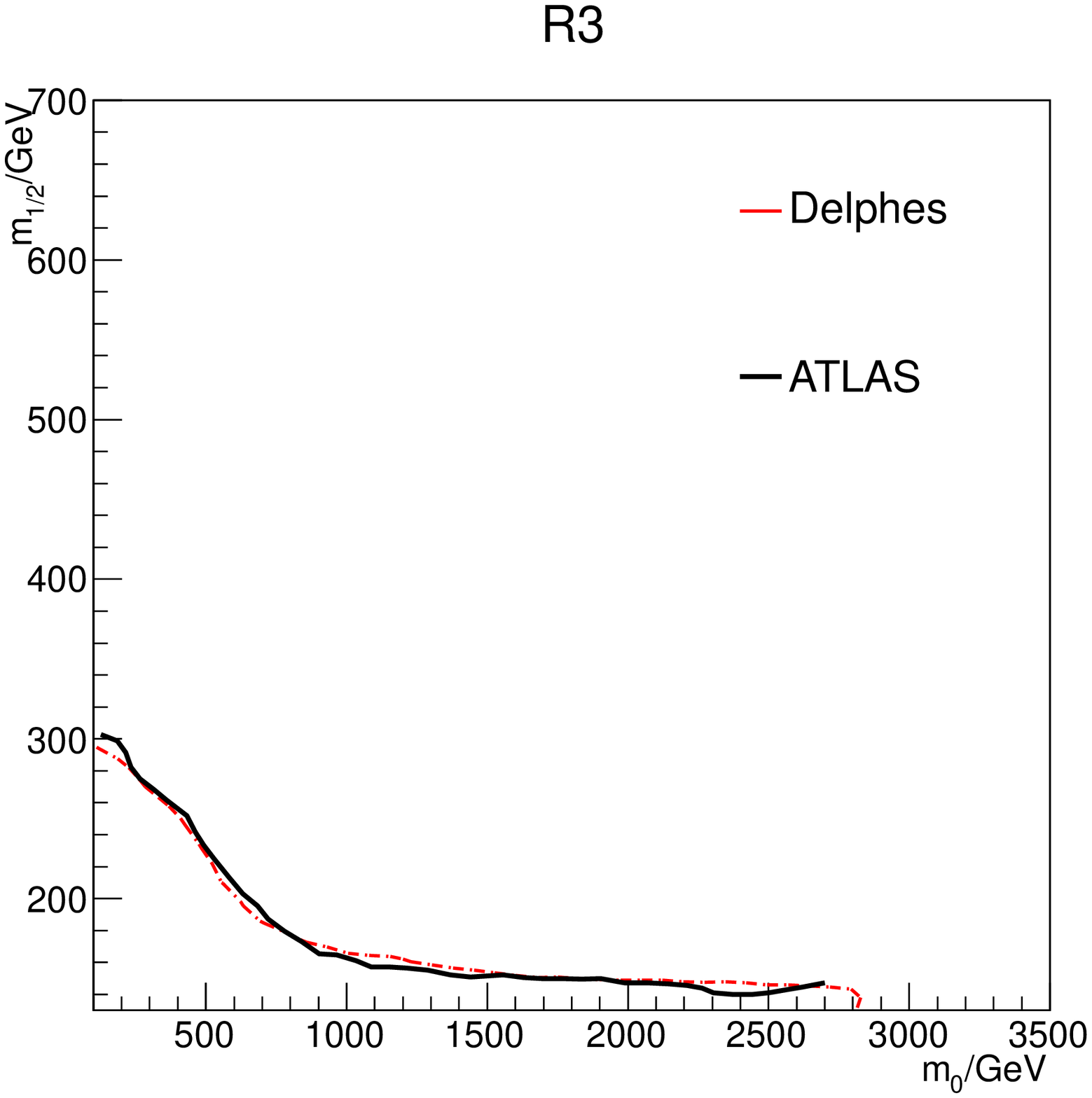}{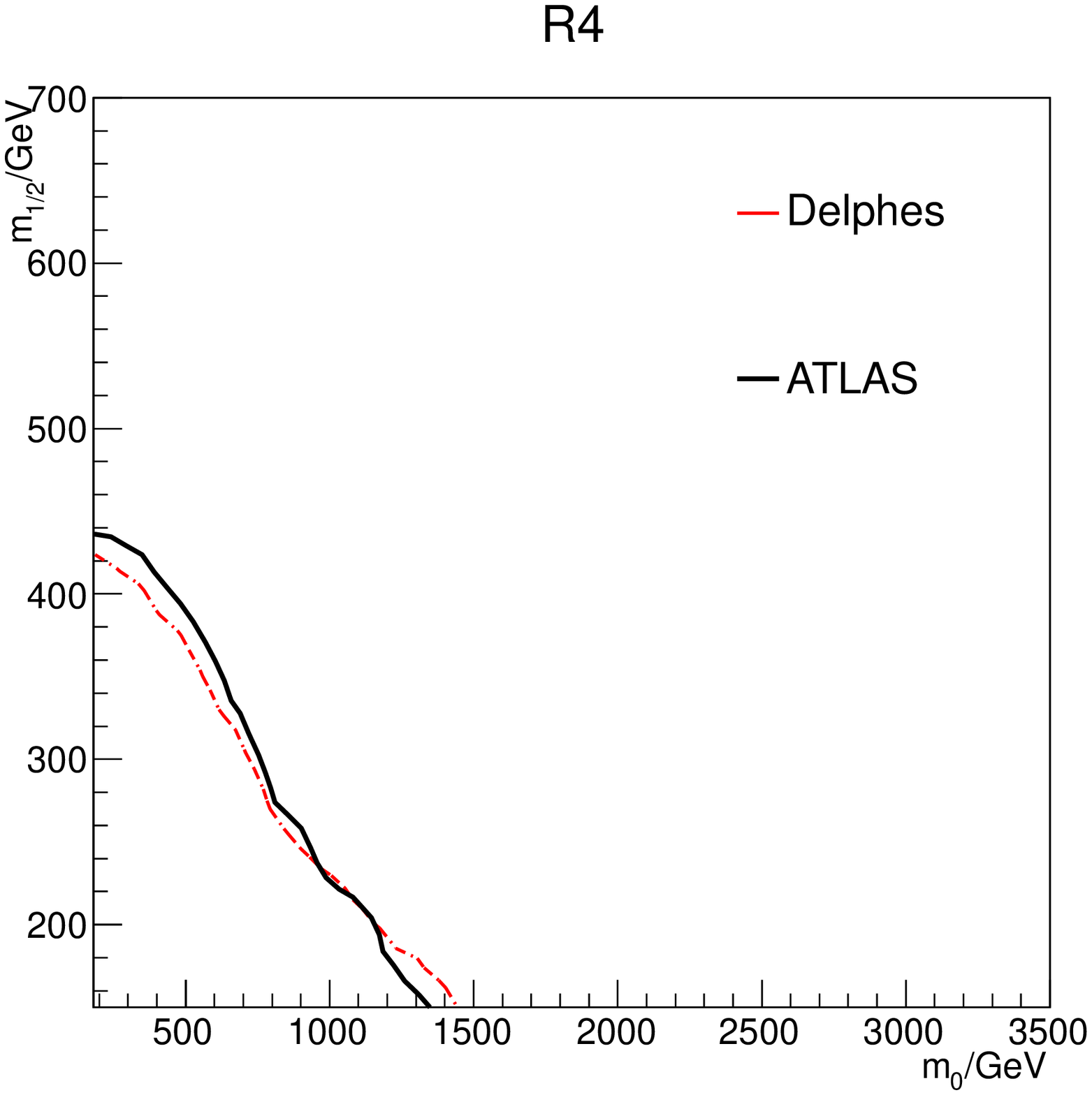}{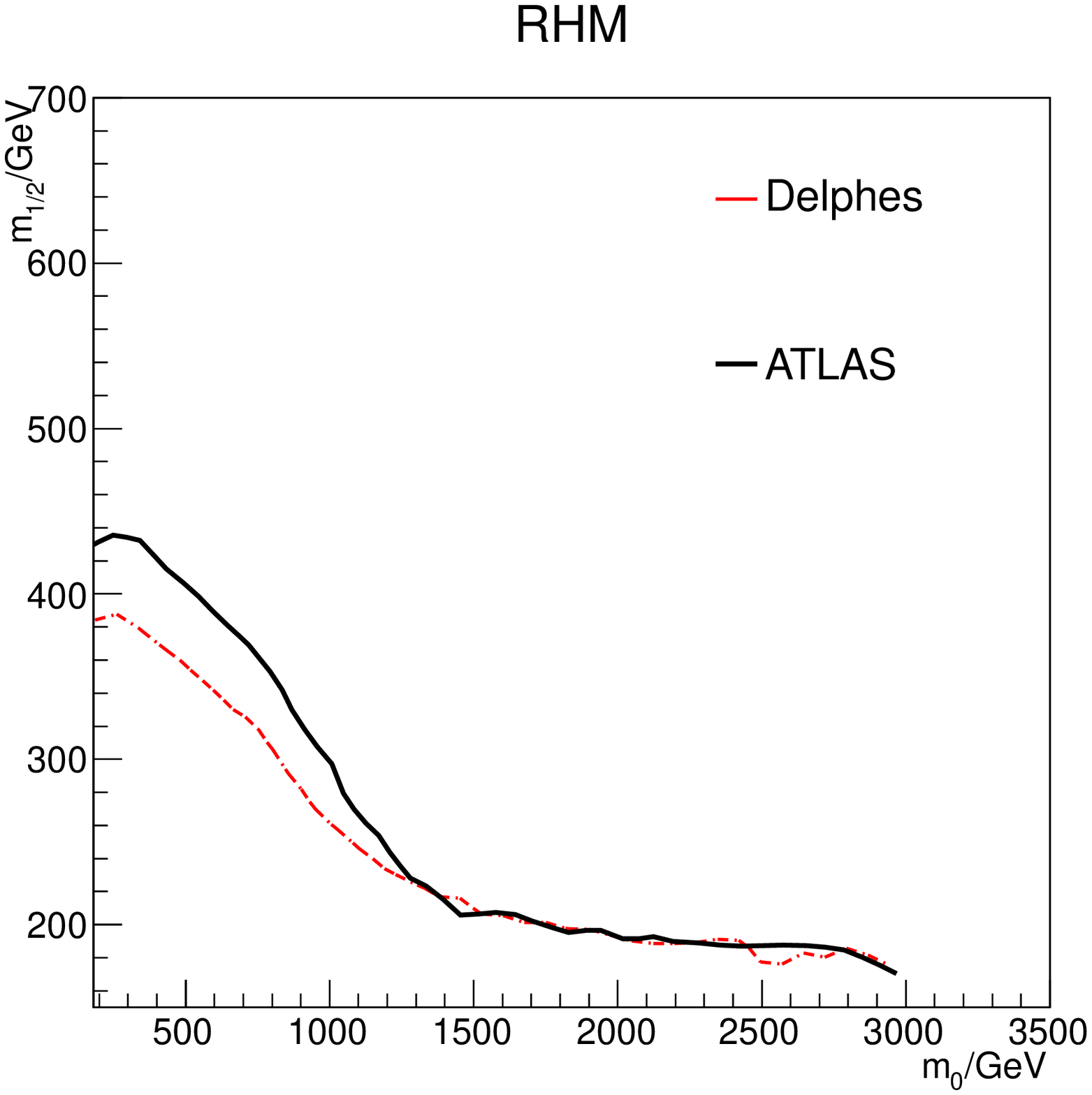}{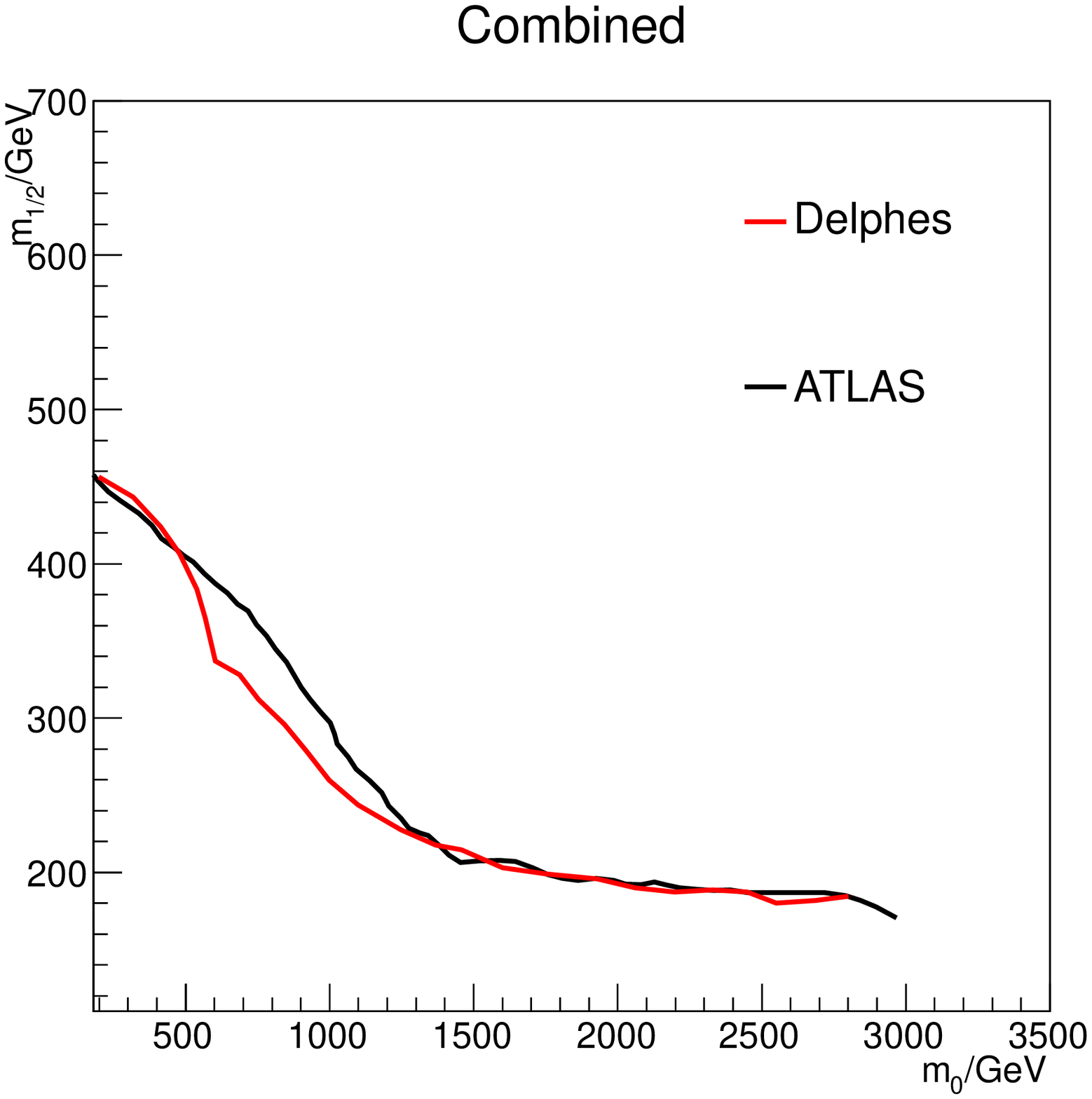}

\caption{Comparison between \tt Delphes \rm and ATLAS 95\% exclusion limits in the $m_0-m_{1/2}$ plane, for the signal regions R1, R2, R3, R4 and RHM defined in  Table~\ref{tab:cuts}. In the combined limit plot, the ATLAS limit is obtained using the ATLAS statistical combination, whilst the \tt Delphes \rm limit is obtained by taking the union of the \tt Delphes \rm limits for each signal region.}
\label{fig:limits1}
\end{figure}

The procedure defined above neglects systematic errors on the signal and background yields and, as noted in~\cite{Allanach:2011qr}, this leads to a discrepancy between the \tt Delphes \rm results and the ATLAS limits in each channel. We follow~\cite{Allanach:2011qr} in using a channel dependent scaling to tune the \tt Delphes \rm output so that the limits in each channel match as closely as possible ``by eye''. We obtain factors of 0.82, 0.85, 1.25, 1.0 and 0.70 for the R1, R2, R3, R4 and RHM regions respectively. Comparisons between the resulting \tt Delphes \rm exclusion limit and the ATLAS limit are shown in Figure~\ref{fig:limits1}, where we observe generally good agreement in all channels. The largest discrepancy is observed in the RHM channel, where we find that one cannot get the tail of the limit at large $m_0$ to agree with the ATLAS limit whilst simultaneously guaranteeing good agreement at low $m_0$. This is likely to be due to the fact that we have effectively assumed a flat systematic error over the $m_0-m_{1/2}$ plane. whereas the ATLAS results use a full calculation of the systematic errors for each signal point. It is important to notice however that the \emph{combined} limit will be dominated by regions R1 and R2 at low $m_0$, and thus by choosing to tune the RHM results in order to reproduce the large $m_0$ tail, one can ensure reasonable agreement of the combined limit over the entire range. Where disagreement remains, the \tt Delphes \rm limit is less stringent than the ATLAS limit, and hence using it gives us a conservative estimate of the ATLAS exclusion reach.

\subsection{Limits on stochastic SUSY space}
Having obtained reasonable agreement with the ATLAS CMSSM limits in the previous section, we now turn our attention to calculating the LHC exclusion zone in the plane of stochastic SUSY parameters $\xi$ and $\Lambda$. To avoid unnecessary (and computationally expensive) simulation, we only simulate those points that passed the constraints detailed in Sec. \ref{sec:scan}. 

\begin{figure}
\begin{center}
\fourgraphs{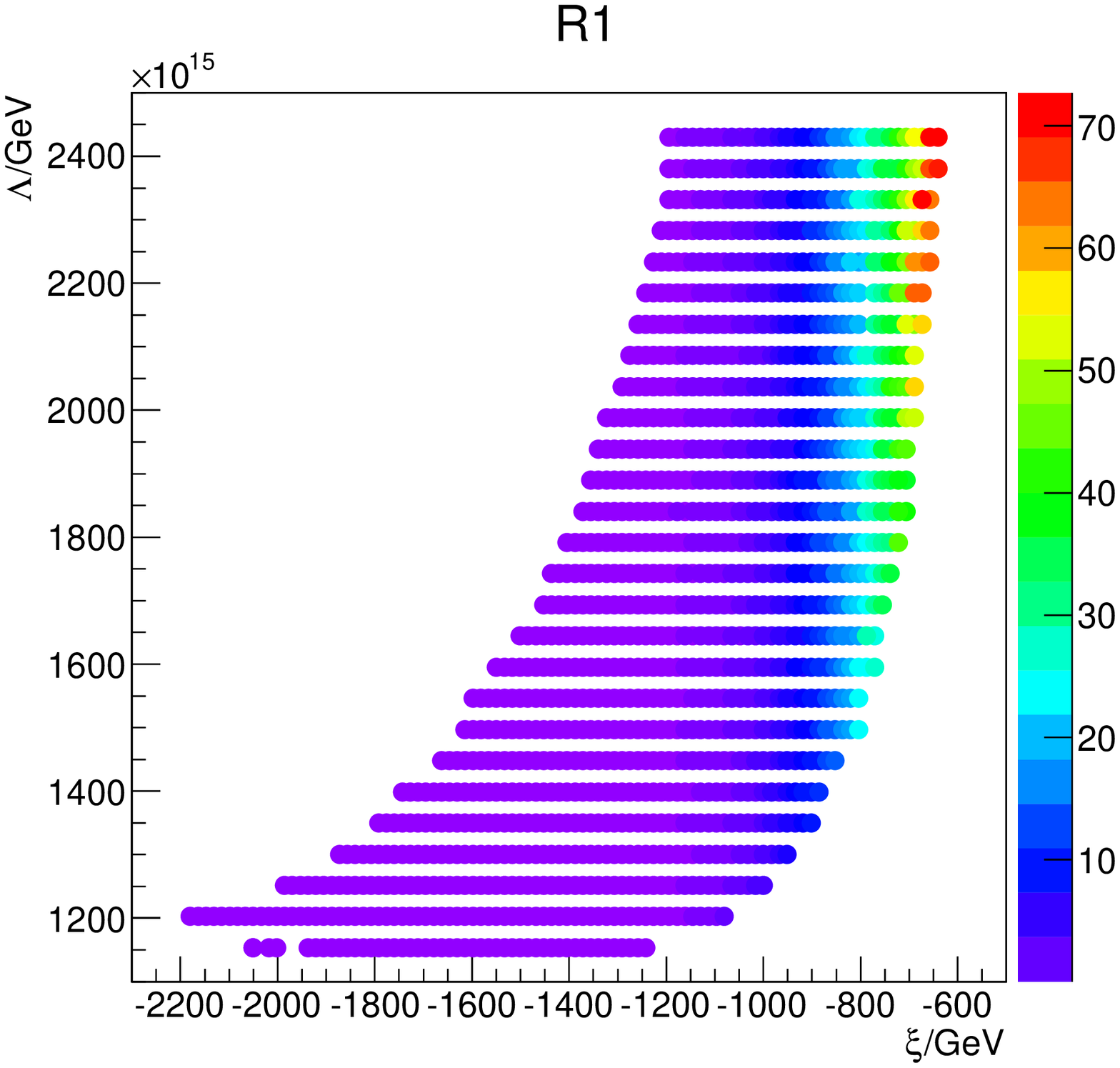}{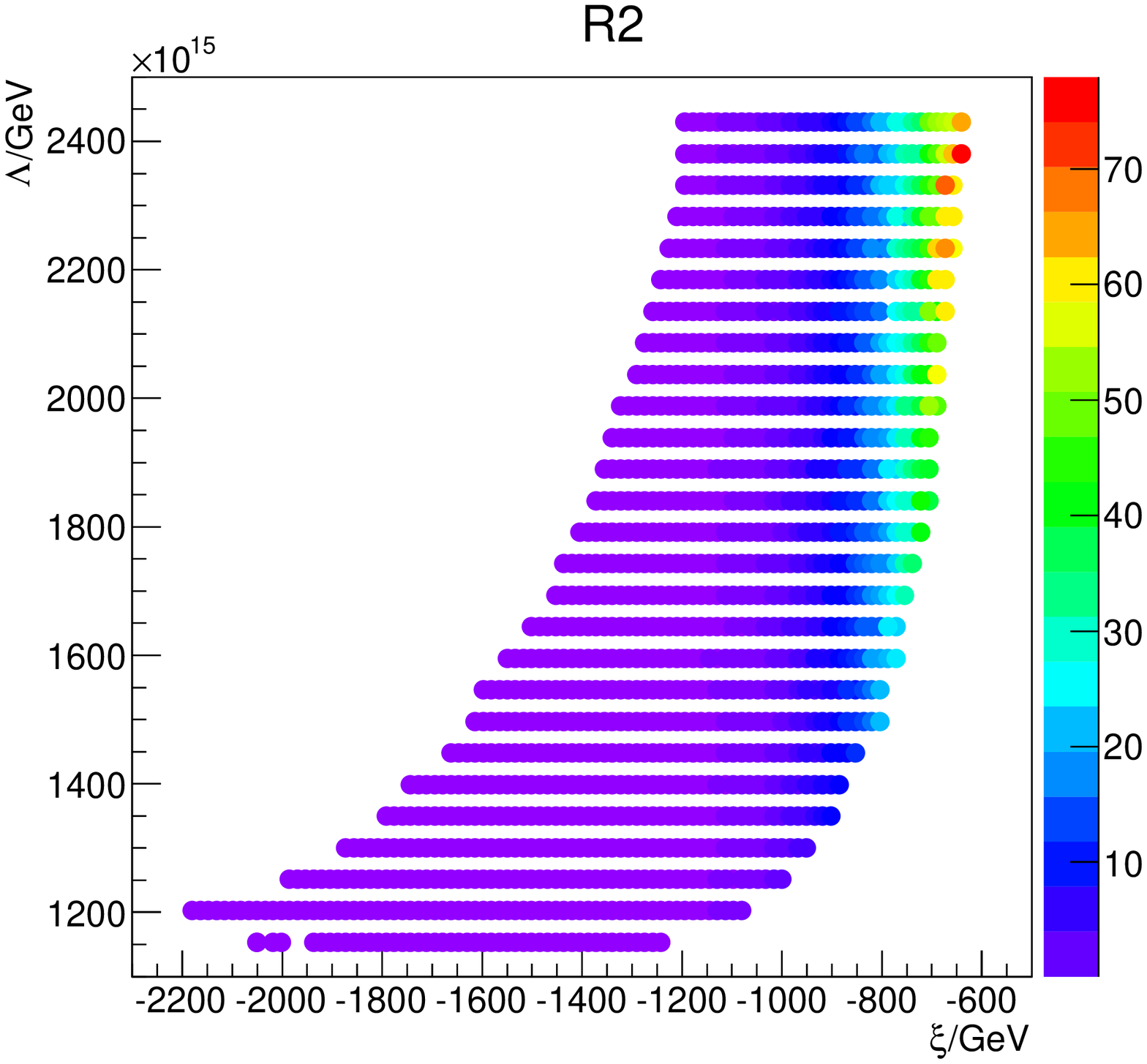}{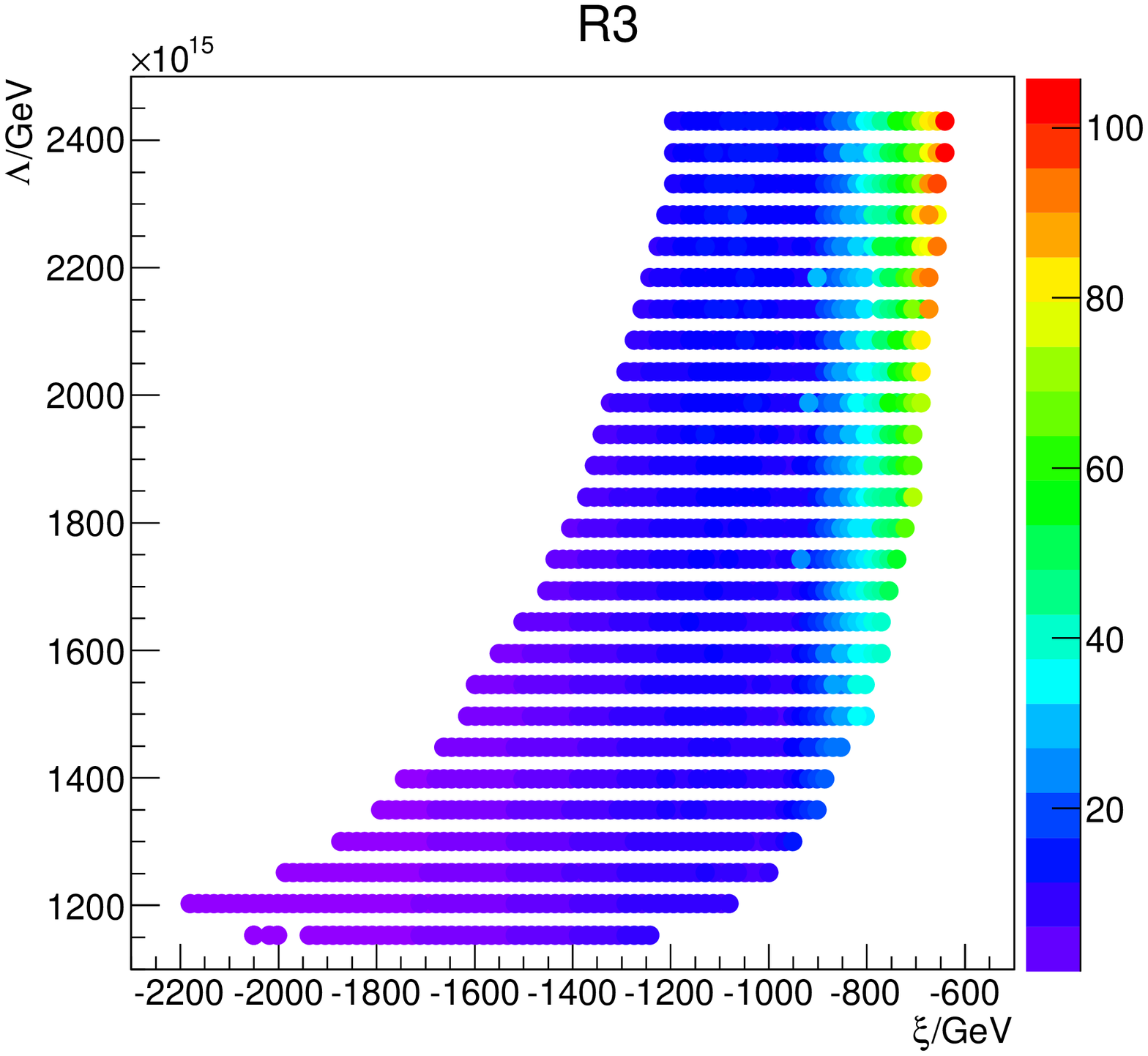}{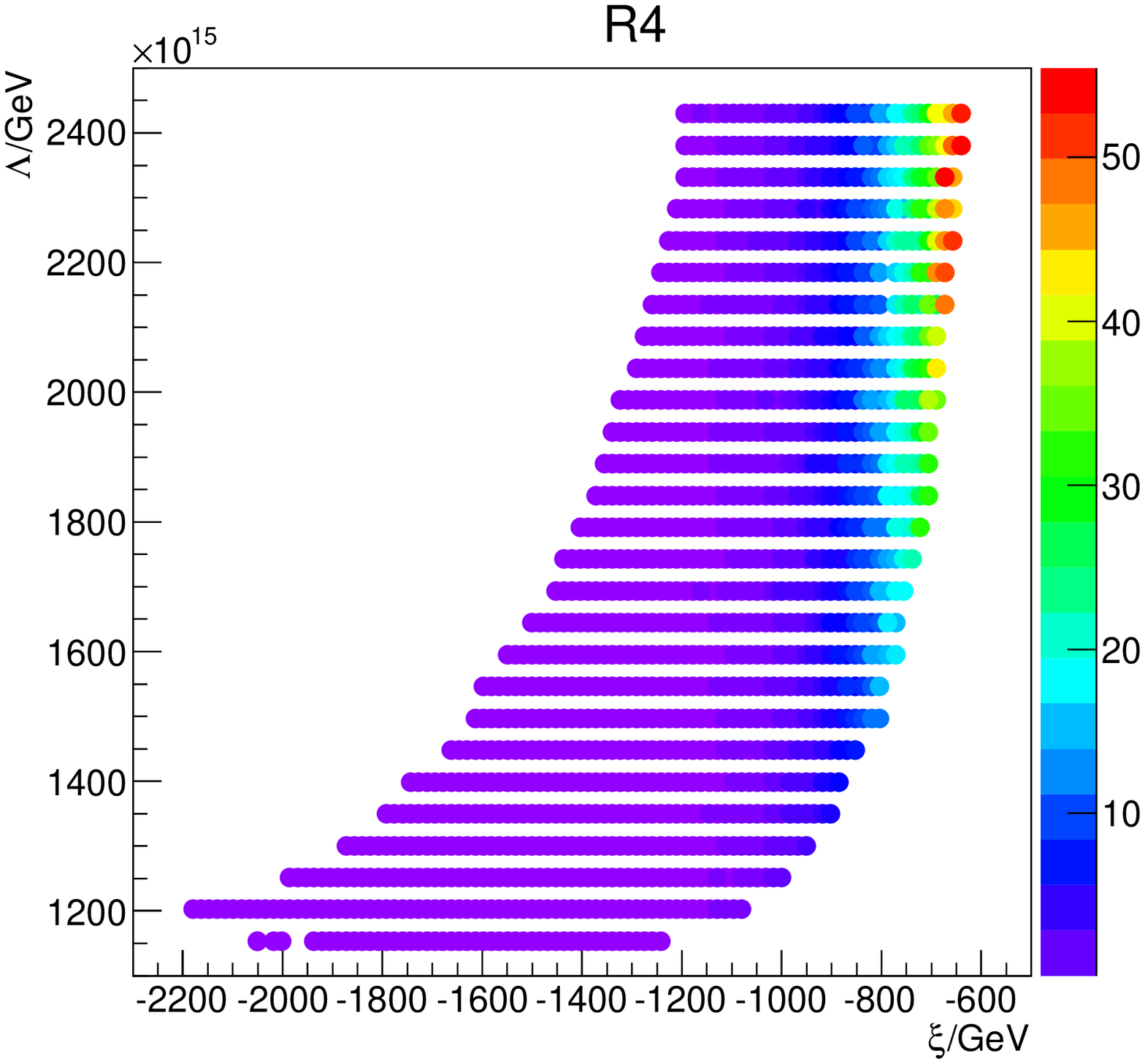}
\caption{$\sigma \times A \times \epsilon$ values for stochastic SUSY points, as obtained using the simulation code detailed in the text, for ATLAS signal regions R1, R2, R3 and R4. The acceptable region of stochastic supersymmetry parameter space can be found by referring to the ATLAS limits shown in Table \ref{tab:results}. Points on the graphs above with values exceeding those limits are excluded. For example, the limit for signal region R1 is 22 fb. All stochastic superspace points with $\sigma \times A \times \epsilon$ larger than 22 fb are ruled out, in this case $\xi \gtrsim -850 \GeV$, represented by the region to the right of the light blue colour.  Correspondingly, the limit for R2 is 25 fb, excluding $\xi \gtrsim -800 \GeV$ in the region to the right of the light blue colour. The limit for R3 is 429 fb, which does not exclude any stochastic superspace points. R4 has a limit of 27 fb, which excludes $\xi \gtrsim -750 \GeV$ in the region to the right of the green coloured squares. }
\label{fig:colz1}
\end{center}
\end{figure}

\begin{figure}
\begin{center}
  \includegraphics[width=\wth]{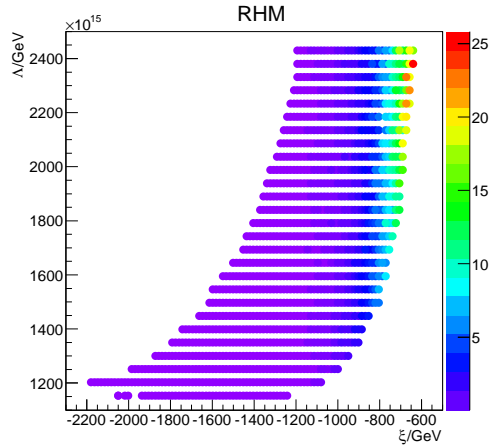}
\caption{$\sigma \times A \times \epsilon$ values for stochastic SUSY points for ATLAS signal region RHM. The ATLAS limit for this signal region is 17 fb, which excludes stochastic superspace points with $\xi \gtrsim -700 \GeV$ in the region to the right of the light green colour.}
\label{fig:colz2}
\end{center}
\end{figure}

Figures~\ref{fig:colz1} and~\ref{fig:colz2} show the $\sigma \times A \times \epsilon$ values (including the systematic factors) for each of the ATLAS search channels. $\sigma \times A \times \epsilon$ is most strongly dependent on $\xi$ which is to be expected; points at low $|\xi|$ have light coloured sparticles and will thus have large production cross-sections at the LHC. Raising the mass scale essentially reduces the production cross-section leading to a corresponding decrease in the $\sigma \times A \times \epsilon$ value as one travels left along the $\xi$ axis. There is little to no dependence of $\sigma \times A \times \epsilon$ on the cutoff scale for a given value of $\xi$. We can conservatively bound the region $\xi \lesssim -900 \GeV$ as consistent with 1 fb$^{-1}$ of LHC data based on ATLAS signal region R1, which is the most constraining in the case of stochastic superspace.

\section{Conclusion}
\label{sec:conclusion}
In this paper, we perform a systematic scan of the parameter space of the stochastic superspace model of soft supersymmetry breaking. Its phenomenology is checked against constraints from direct sparticle searches, cold dark matter relic density and the branching ratio of the process $B_s \rightarrow \mu^+ \mu^-$. Points that pass these conditions are analysed further, obtaining a value for $\sigma \times A \times \epsilon$, which can be directly compared to ATLAS search limits from the zero lepton channels. We find that the stochasticity parameter is restricted to the region $\xi \lesssim -900 \GeV$, while the viable region of the cutoff scale, $\Lambda$, shrinks with increased $\left| \xi \right|$, bounded above by relic density concerns, and below from sparticle mass constraints. 

Since a prime motivation for the the low-energy supersymmetry is a solution to the hierarchy problem, the increased bounds on sparticle masses seriously undermine its validity. The required tuning of parameters is estimated at the level of $\sim 10\%$ for $\xi \sim -1000 \GeV$ \cite{Kobakhidze:2010ee}. Thus, stochastic supersymmetry still allows a region of parameters that provide a satisfactory solution to the hierarchy problem. However, as has been shown in \cite{Kobakhidze:2008py,Kobakhidze:2010ee}, the mass of the lightest Standard Model-like Higgs boson is predicted in a very narrow range $m_{\rm h} \approx 112-116 \GeV$ in the minimal model. This range is outside the $124-126 \GeV$ region for the Higgs mass hinted in the very recent ATLAS \cite{ATLAS-CONF-2011-163} and CMS \cite{CMS-PAS-HIG-11-032} data. If the latest evidence for the Higgs mass will be confirmed, the minimal model with stochastic supersymmetry will certainly be excluded, motivating research into a non-minimal implementation of stochastic superspace consistent with these results.

\acknowledgments
This work was supported in part by the Australian Research Council.

%\bibliography{test}

\end{document}